\documentclass[prd,10pt,aps, twocolumn,nofootinbib,showpacs,floatfix,amssymb,amsmath,preprintnumbers,superscriptaddress]{revtex4-1}
\usepackage{rotating,slashed,amssymb}

\newcommand{\beqn}{\begin{eqnarray}}
\newcommand{\eeqn}{\end{eqnarray}}
\newcommand{\eq}[1]{(\ref{#1})}
\newcommand{\cL}{{\cal L}}

\newcommand{\cP}{{\cal P}}
\newcommand{\cO}{{\cal O}}

\newcommand{\cA}{{\cal A}}
\newcommand{\cD}{{\cal D}}

\newcommand{\tr}{ {\rm Tr} \, }

\newcommand{\Dirac}{\slashed \cD}

\newcommand{\lr}[1]{ \left( #1 \right) }

\newcommand{\vev}[1]{ \langle \, #1 \, \rangle }

\begin{document}

\title{Temperature dependence of the axial magnetic effect in two-color quenched QCD}

  \author{V. Braguta}
   \affiliation{IHEP, Protvino, Moscow region, 142284 Russia}
   \affiliation{ITEP, B. Cheremushkinskaya str. 25, Moscow, 117218 Russia}

  \author{M. N. Chernodub}\email{On leave from ITEP, Moscow, Russia.}
\affiliation{CNRS, Laboratoire de Math\'ematiques et Physique Th\'eorique,
Universit\'e Fran\c{c}ois-Rabelais Tours,\\ F\'ed\'eration Denis Poisson, Parc
de Grandmont, 37200 Tours, France}
\affiliation{Department of Physics and Astronomy, University of Gent, Krijgslaan
281, S9, B-9000 Gent, Belgium}

\author{V. A. Goy}
\affiliation{School of Natural Sciences, Far Eastern Federal University, Sukhanova str., 8, Vladivostok, 690950, Russia}

\author{K. Landsteiner}
     \affiliation{Instituto de F\'{\i}sica Te\'orica UAM/CSIC, C/ Nicol\'as
Cabrera 13-15,\\
     Universidad Aut\'onoma de Madrid, Cantoblanco, 28049 Madrid, Spain} 

\author{A. V. Molochkov}
\affiliation{School of Biomedicine, Far Eastern Federal University, Sukhanova str., 8, Vladivostok, 690950, Russia}

\author{M. I. Polikarpov}\email{Deceased.}
     \affiliation{ITEP, B. Cheremushkinskaya str. 25, Moscow, 117218 Russia}
     \affiliation{Moscow Inst Phys \& Technol, Institutskii per. 9, Dolgoprudny,
Moscow Region, 141700
Russia}

\date{January 31, 2013}
 
\begin{abstract}
The Axial Magnetic Effect is the generation of an equilibrium dissipationless energy
flow of chiral fermions in the direction of the axial (chiral) magnetic field.
At finite temperature the dissipationless energy transfer may be realized in the
absence of any chemical potentials.  We numerically study the temperature
behavior of the Axial Magnetic Effect in quenched $SU(2)$ lattice gauge theory.
We show that in the confinement (hadron) phase the effect is absent. In the
deconfinement transition region the conductivity quickly increases, reaching the
asymptotic $T^2$ behavior in a deep deconfinement (plasma) phase. Apart from an
overall proportionality factor, our results qualitatively agree with theoretical
predictions for the behavior of the energy flow as a function of temperature and
strength of the axial magnetic field. 
\end{abstract}
 \preprint{IFT-UAM/CSIC-14-002}
 \maketitle

One of the cornerstones of modern quantum field theory is the concept of 
anomalies. A symmetry present at the classical level gets broken by
the effects of quantum mechanics. Anomalies are responsible for quantum
processes that would not occur in their absence, e.g. the decay of the
neutral pion into two photons. 

In the recent years it has become increasingly clear that anomalies have also
important 
consequences in the transport
properties of a gas or liquid whose constituents have chiral fermions amongst
them.
These effects consist of the generation of dissipationless currents in the
presence
of a magnetic field or a vortex. They are called the Chiral Magnetic Effect \cite{ref:CME}
and the Chiral Vortical Effect \cite{Erdmenger:2008rm}. The currents can either
be global (anomalous) currents 
such as the axial current or (conserved) gauge currents such as the electric
current or the energy current. 
These transport phenomena are conveniently described by a set of transport
coefficients, the
chiral conductivities. They depend on the chemical potentials and the
temperature. While
the dependence on the chemical potentials is related to the conventional chiral
anomalies
the temperature dependence enters via the gravitational contribution to the
anomalies \cite{Landsteiner:2011cp}. 
On the level of Feynman diagrams the conventional anomalies appear in triangles
diagrams of currents
whereas the gravitational anomaly appears in  triangles with one current and two
energy
momentum tensors. 

A form of these transport laws has been derived many years ago in the context of
neutrino physics \cite{Vilenkin:1979ui}.
Their universal character and the deeper relation to anomalies have been
realized only recently. The 
relation to anomalies is most striking in the framework of hydrodynamics. It was
shown that the
hydrodynamic constitutive relations for an anomalous current necessarily have to
include the chiral
magnetic and chiral vortical effects and that the dependence of the chiral
conductivities on the
chemical potentials are completely
fixed with this framework \cite{Son:2009tf}. The temperature dependence is fixed
via a combination
of hydrodynamic and geometric reasoning \cite{Jensen:2012kj}. In essence this
result constitute 
non-renormalization theorems for the chiral conductivities. 

However, in Ref.~\cite{Golkar:2012kb} it was shown that gauge interactions do give
rise to a non-vanishing two loop contribution
to the temperature dependence of the chiral conductivities. In our previous
(limited) lattice study \cite{Braguta:2013loa}
a large suppression of the temperature dependence compared to the weak coupling
result was found.
Further higher loop corrections to chiral conductivities have been shown to
appear in Ref.~\cite{Jensen:2013vta}.

In all these cases dynamical gauge fields are
present. We can distinguish anomalies as ``quantum'' or
``classical'' whether the divergence of the current is given by an expression
containing only
classical fields or by a quantum operator. More explicitly, in the anomalous
non-conservation law
$\partial_\mu J^\mu = c F\tilde F$, $F$ might be the field strength of a purely
external, non-dynamical
field $A_\mu$ whose purpose is to act as a source for the quantum operator
$J^\mu$ in the effective
action. In this case we can speak of a c-number anomaly. The current $J$ can in
this case be incorporated
in a hydrodynamic formulation and the non-renormalization theorems of
\cite{Son:2009tf, Jensen:2012kj} apply.
If we are interested however in the axial current in QCD the field strength
appearing in the anomaly equation
is the gluonic one and we also need to take the quantum dynamics of the gluon
fields into account. In that
case the anomaly is a q-number, i.e. an operator. For the axial current in QCD
it is given by the topological 
charge density. Now the axial charge is bound to undergo quantum
fluctuations\footnote{It is
precisely these quantum fluctuations that are thought to be responsible for the
chiral magnetic effect in
heavy ion collisions~\cite{ref:CME}. }. In this case the values of the anomalous
conductivities can and do
suffer renormalization from interactions via dynamical gauge fields. From these
considerations it becomes 
clear that in order to understand the role anomalous transport plays
in heavy ion collisions it is of utmost importance to know how much the values
of the anomalous conductivities can
be modified in the strong coupling regime.

Anomalous conductivities
describe the presence of dissipationless equilibrium currents. Therefore they
are in principle accessible to inherently
Euclidean lattice gauge theory. Most of the chiral conductivities do depend
however crucially on chemical potentials
whose lattice implementation is notoriously difficult. Luckily there is however
one chiral conductivity that is non-vanishing
even at zero chemical potential if only the system is at finite temperature.
This is the chiral vortical conductivity
in the axial current
\begin{equation}
 \label{eq:cveaxial}
\vec{J}_5 = \sigma_{\mathrm{CVE},5} \,\vec\omega\,,
\end{equation}
where $\vec\omega = \vec\nabla \times \vec v$ is the vorticity. This equation
would still be difficult to
handle on the lattice since it describes the response of the system to rotation.
Another, closely related effect is the
so-called Axial Magnetic Effect (AME). It describes the generation of an energy
current $J^i_\epsilon = T^{0i}$
in the background of an
axial magnetic field, i.e. a magnetic field that couples with opposite signs to
left-handed and right-handed fermions,
\begin{equation}
 \label{eq:ame}
 \vec{J}_\epsilon = \sigma_{\mathrm{AME}}\, \vec{B}_5\,.
\end{equation}
The Kubo formulae for these two chiral conductivities imply
$\sigma_{\mathrm{CVE},5} = \sigma_{\mathrm{AME}}$.
At weak coupling and for a collection of left-handed fermions with charges $q_L$
and 
right-handed fermions with charges $q_R$ the axial magnetic conductivity is
given by
\begin{equation}\label{eq:sigma}
 \sigma_{\mathrm{AME}} = \frac{1}{24} \left(\sum_R q_R - \sum_L q_L \right)
T^2\,.
\end{equation}

It is relatively simple to check the AME transport law~\eq{eq:ame} in Euclidean
lattice simulations because the AME may be realized in the pure thermal vacuum
with all chemical potentials set to zero, $\mu = \mu_5 = 0$. On the other hand,
the lattice implementation of the axial magnetic field ${\vec B}_5$ is
a straightforward procedure~\cite{Braguta:2013loa,Buividovich:2013hza}.

According to Eq.~\eq{eq:ame} the axial magnetic field should induce a
dissipationless energy flow of the quarks along the axis of the field.  In our
previous paper we have confirmed the emergence of the Axial Magnetic Effect
using lattice simulations in quenched SU(2) QCD~\cite{Braguta:2013loa} for a
certain temperature in the deconfinement phase.
The same effect has been demonstrated for a system of free lattice
fermions~\cite{Buividovich:2013hza}.

In the deconfinement phase the energy flow turned out to be proportional to the
strength of the axial magnetic field in a qualitative agreement with the
analytical prediction~\eq{eq:ame}. Theoretically, the transport AME
law~\eq{eq:ame} was derived in a linear response theory so that Eq.~\eq{eq:ame}
should in principle be valid only in a weak field limit. Surprisingly, the
numerical results of Ref.~\cite{Braguta:2013loa} have shown that the linear
behavior in $B_5$ persists up to very high magnetic fields $e B_5 \approx 1.2\,
{\mathrm{GeV}}^2$. The simulations confirm with a high accuracy the linear
behavior of the AME law~\eq{eq:ame} in the whole range of studied axial magnetic
fields. 

In the confinement phase $T < T_c$, and in the region close to the phase
transition, $T \sim T_c$, the dissipationless energy transfer ceases to exist.
The disappearance of the effect in the low-temperature region is a natural
consequence of the fact that the AME law is essentially based on properties of
the quarks, which are absent in the spectrum of the theory at low temperatures
due to the quark confinement phenomenon. 

In order to illustrate the existence of the effect it was sufficient to consider
one fermion species, $N_f = 1$, with a unit charge:
\beqn
q^L_{5} = - q^R_{5} = + e\,.
\label{eq:RL:charges}
\eeqn
According to Eq.~\eq{eq:sigma}, in QCD with two colors $N_c = 2$ and one flavor
$N_f = 1$ the prefactor $\sigma$ in the AME law~\eq{eq:ame} has the following
form:
\beqn
\sigma^{\mathrm{th}}(T) = C^{\mathrm{th}}_{\mathrm{AME}} \, T^2\,, \qquad
C^{\mathrm{th}}_{\mathrm{AME}} = \frac{2 N_f N_c}{24} \equiv \frac{1}{6}\,.
\label{eq:sigma:theor}
\eeqn
The temperature behavior~\eq{eq:sigma:theor} of the conductivity coefficient
$\sigma$ is assumed to be valid in a deep deconfinement phase, far from the
low-temperature confining region. 

In Ref.~\cite{Braguta:2013loa} the AME law~\eq{eq:ame} was studied numerically
at the single temperature $T
= 480\, {\mathrm{MeV}} \simeq 1.58 T_c$ in the deconfinement phase where 
$T_c = 303\,\mathrm{MeV}$ is the critical temperature of the deconfinement phase
transition
of the lattice $SU(2)$ gauge theory in the continuum
limit~\cite{Fingberg:1992ju}. 

We have found that the
proportionality coefficient $C_{\mathrm{AME}}$ in the AME conductivity
$\sigma(T) = C_{\mathrm{AME}}\, T^2$ substantially differs from the theoretical
prediction \eq{eq:sigma:theor}, 
\beqn
\frac{C_{\mathrm{AME}}(T)}{C^{\mathrm{th}}_{\mathrm{AME}}} {\Biggr |}_{T = 480
\,{\mathrm{MeV}}} \simeq 0.058\,.
\label{eq:ratio}
\eeqn
The large difference between the theoretical and numerical results~\eq{eq:ratio}
could be a result of a peculiar temperature behavior of the proportionality
coefficient $C_{\mathrm{AME}}(T)$, so that the asymptotic
regime~\eq{eq:sigma:theor} may not have been reached at the studied temperature
$T = 480\, {\mathrm{MeV}}$.

The aim of this paper is to find the temperature behavior of the dissipationless
energy transport~\eq{eq:ame} in a wide range of temperatures. In particular, we
are interested in confirmation of the $T^2$ behavior of the proportionality
coefficient predicted by the theory and verification of the validity of the
theoretical prediction~\eq{eq:sigma:theor} in a deep deconfinement phase. To
this end, we extend the calculations of Ref.~\cite{Braguta:2013loa} to a larger
set of temperatures and increase statistics of numerical simulations. Below we
briefly overview our numerical techniques for the sake of completeness.

We consider the following Lagrangian,
\beqn
\cL_5 = {\bar \psi} (\partial_\mu - i g A^a_\mu t^a - i \gamma^5 e A_{5,\mu})
\gamma^\mu  \psi \equiv {\bar \psi} \Dirac_5 (A_5)\psi\,, \quad
\label{eq:D5f}
\eeqn
where the axial field $A_{5,\mu}$ acts as a classical background field
superimposed over the dynamical non-Abelian field $A^a_\mu$ is a gauge field. In
Eq.~\eq{eq:D5f} $t^a$, $a = 1,2,3$ are the generators of the corresponding
$SU(2)$ gauge group. The gauge field $A^a_\mu$ is generated in lattice
Monte-Carlo simulations of $SU(2)$ lattice gauge theory. 

We are working in the quenched approach so that a backreaction of the fermions
on the non-Abelian gauge field via the vacuum fermion loops is disregarded. It is 
known that the quark propagator is not strongly altered by the quenching effects~\cite{ref:quenching}
therefore one may generally expect that the quenching should give a moderate contribution to the 
anomalous transport of quarks in Eq.~\eq{eq:sigma:theor}. The reduced number of colors 
(2 instead of 3) is already taken into account in the theoretical estimate~\eq{eq:sigma:theor}.

In our simulations we choose the axial gauge field in the following form 
\beqn
A_{5,0} = A_{5,3} = 0, \quad A_{5,1} = - \frac{x_2 B_5}{2}, \quad A_{5,2}  =
\frac{x_1 B_5}{2}, \quad
\label{eq:A:choice}
\eeqn
which corresponds to a stationary uniform axial magnetic field pointing in the
third direction, $B_{5,i} = B_5 \cdot \delta_{i,3}$. Here the latin index $i =
1,2,3$ labels the spatial coordinates and $\mu=0$ is the time direction. The
axial electric field is absent. 

According to Eq.~\eq{eq:ame} the axial magnetic background~\eq{eq:A:choice}
should induce the dissipationless energy flow of the quarks. The latter is given
by the expectation value of the $T^{0i}$ component of the stress-energy tensor,
\beqn
 J^i_\epsilon & = & \vev{T^{0i}} \equiv \frac{i}{2} \vev{{\bar \psi}  (\gamma^0
\cD^{i}_5 + \gamma^i \cD^{0}_5)\psi }\,.
\label{eq:J:E}
\eeqn
In addition to the fermionic part~\eq{eq:J:E}, the energy flow should also contain 
a gluonic contribution. Although the gluons carry no electric charge, their dynamics 
is affected by the external magnetic field via interactions with quark vacuum loops. 
However, in the quenched approach the quark vacuum loops are absent so that 
the gluons are not sensitive to the external magnetic field. As a result the energy 
flow of gluons is vanishing in our approach.

The lattice implementation of the continuum formula~\eq{eq:J:E} is achieved via
a straightforward discretization,
\beqn
C_\mu (x,y;A) & = & \vev{\bar{\psi}  (x) U_{x,y}(A^a_\mu) \gamma_{\mu} {\psi}
(y)}_A \nonumber \\
& \equiv & \tr\lr{U_{x,y}(A^a_\mu) \, \frac{1}{\Dirac_5 + m} \,
\gamma_{\mu}}_{x,y;A}\,,
\label{eq:core:latt}
\eeqn
where the expectation value is taken over the fermion field in a fixed
background of axial $A_{5,\mu}$ and non-Abelian $A^a_\mu$ fields. These fields
enter the Dirac operator $\cD_5$ which is defined in Eq.~\eq{eq:D5f}. In
Eq.~\eq{eq:core:latt} trace operation is taken over color and spinor indices,
and $U_{x,y}$ is the gluon string between the lattice points $x$ and $y$ which
makes Eq.~\eq{eq:core:latt} gauge invariant. The expectation
value~\eq{eq:core:latt} should eventually be averaged over the ensemble of the
dynamical gauge fields $A^a_\mu$. 

The fermion propagator~\eq{eq:core:latt} is calculated using the following
identity,
\beqn
& & {\mathrm{tr}} \left[S_5(A_5) \, \gamma_{\mu}\right] \equiv {\mathrm{tr}}
\left[ (\cP_R + \cP_L) S_5(A_5) \, \gamma_{\mu}\right] \nonumber \\
& & \quad\, = {\mathrm{tr}} \left[\cP_R \, S(A_5) \gamma_{\mu}\right] +
{\mathrm{tr}} \left[\cP_L \, S(-A_5) \gamma_{\mu}\right], \qquad
\label{eq:identity}
\eeqn
where the trace is taken over spinor indices and $\cP_{R,L} = (1 \pm
\gamma^5)/2$ are the right and left chiral projectors, respectively. The
identity~\eq{eq:identity} expresses the trace of the propagator $S_5(A_5)$ in a
background of the axial field $A_5$ via the traces of the usual propagators 
$S(\cA)$ calculated in the background of the standard $U(1)$ gauge fields
$\cA$, 
\beqn
S_5(A_5) =  \left[\Dirac_5 (A_5)\right]^{-1}, \qquad S(\cA) & = & \left[\Dirac
(\cA)\right]^{-1}.
\eeqn
The Dirac operator for the former is defined in Eq.~\eq{eq:D5f} while the one
for the latter has the usual form:
\beqn
\cD_\mu(\cA) = \partial_\mu - i g A^a_\mu t^a - i e \cA_\mu\,.
\eeqn
In Eq.~\eq{eq:identity} the axial gauge field $A_5$ appears as the Abelian field
coupled with opposite signs to right-handed and left-handed fermions, in
agreement with the prescription for the left- and right-handed
charges~\eq{eq:RL:charges}.

The correlation functions~\eq{eq:core:latt} are calculated according to the
numerical setup of Refs.~\cite{Buividovich:2009wi}. The quark fields are
simulated with the help of the overlap lattice Dirac operator $\mathcal{D}$ with
exact chiral symmetry~\cite{Neuberger:98:1}. The discretized version of
Eq.~\eq{eq:J:E} is calculated using the correlation functions~\eq{eq:core:latt},
which are averaged over an equilibrium finite-temperature ensemble of
non-Abelian gauge field configurations $\cA_\mu$:
\begin{eqnarray}
\left\langle \cO \right\rangle = \left(\int D A^a_{\mu}\,e^{-S_{YM} [A^a_{\mu}]
}\right)^{\hskip -0.5mm -1} \hskip -1.5mm  \int D A^a_{\mu}\,e^{-S_{\mathrm{YM}}
[A^a_{\mu}] }\, \cO\,,
\quad
\nonumber
\end{eqnarray}
where $S_{\mathrm{YM}}(A^a_{\mu})$ is the Yang-Mills lattice action. 

We evaluate the energy flow~\eq{eq:J:E} using 2700 gauge configurations for
every value of parameters (spatial $L_s$ and temporal $L_t$ lattice sizes,
lattice spacing $a$ and strength of the axial gauge field $eB_5$). In our
previous simulations~\cite{Braguta:2013loa} -- which were carried out with much
smaller statistics -- we have consider the asymmetric lattices $L_s^3 L_t$ with
three temporal lengths $L_t = 4,6,8$ and the fixed spatial length $L_s = 14$. In
addition, we have checked the robustness of the results with respect to
variations of the volume and the lattice spacing. 

In this paper we explore the high-temperature part of the phase diagram
concentrating on single value of the temporal lattice extension $L_t = 4$ and
larger spatial lattice volumes $L_s = 16,\, 18,\, 20$. We make our simulations
for the physical lattice spacings in the interval $a = (0.068 \dots 0.148)$\,fm
and the temperature range $T = (330 \dots 720) \,{\mathrm{MeV}}$.

We use the improved lattice action for the gluon
fields~\cite{Bornyakov:2005iy}. 
Due to the identity~\eq{eq:identity}, the axial magnetic field shares many
properties of the usual magnetic field. For example, the strength of the axial
magnetic field is subjected to  quantization due to the periodicity of the gauge
fields in a finite lattice volume:
\beqn
B_5 = k \, B_{5,\mathrm{min}}\,, \qquad 
e B_{5,\mathrm{min}} = \frac{2 \pi}{L^2_s}\,.
\qquad
\label{eq:B:quant}
\eeqn
Here the integer number $k = 0, 1, \dots, L_s^2/2$ determines the total number
of elementary
magnetic fluxes threading each $(x^1,x^2)$ plane of the lattice. The
quantization~\eq{eq:B:quant} is consistent with the unit charges of the left-
and right-handed quarks~\eq{eq:RL:charges}. In order to avoid ultraviolet
artifacts, we simulate the lattice at relatively small values of the flux quanta
$k \leq 15$ which is much smaller than the maximal possible value of the
quantized flux, $k_{\mathrm{max}} = L^2_s/2 \sim 100$. Our typical strongest
magnetic fields are of the order $e B_{5,\mathrm{max}} \sim 1.\,
{\mathrm{GeV}}^2$ while the smallest possible fields are of the order of $e
B_{5,\mathrm{min}} \sim 0.1\, {\mathrm{GeV}}^2$.

We have numerically checked that the dissipationless energy flow scales linearly
with the strength of the axial magnetic field $B_5$  for a wide set of
temperature and volumes, in agreement with the theoretical
prediction~\eq{eq:ame} and our previous numerical
calculations~\cite{Braguta:2013loa}. Thus, in order to find the temperature
behavior of the conductivity coefficient,
\beqn
C_{\mathrm{AME}}(T) = \frac{J_\epsilon(T, eB_5)}{eB_5 T^2}\,,
\label{eq:coeff}
\eeqn
it is sufficient to calculate the energy current $J_\epsilon$ for a single value
of the external axial magnetic field $B_5$ at a given temperature $T$.

\begin{figure}[!htb]
\begin{center}
\includegraphics[scale=0.575,clip=false]{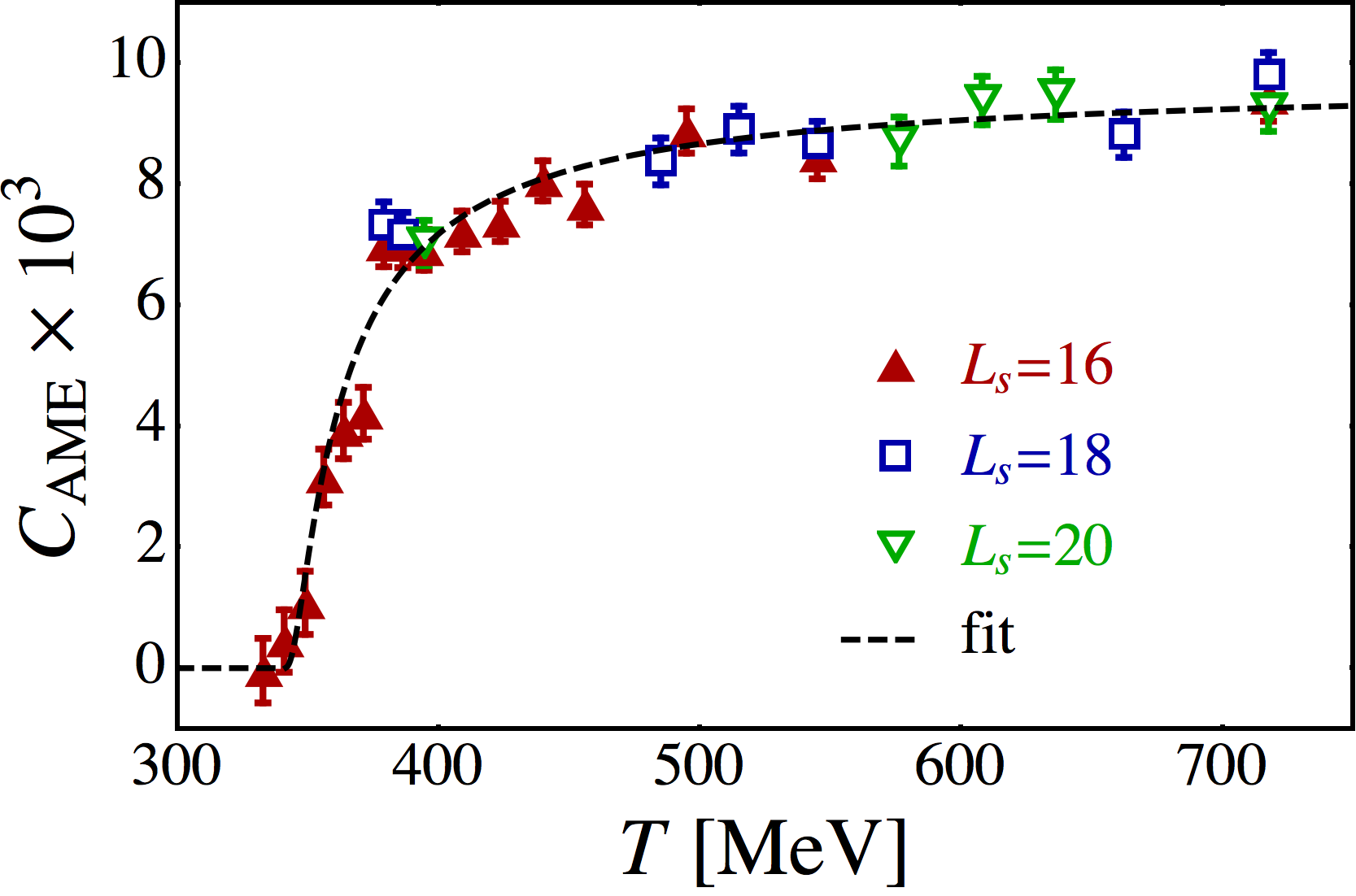} 
\end{center}
\vskip -5mm
\caption{The dimensionless conductivity coefficient~\eq{eq:coeff} of the
dissipationless energy flow vs. temperature. The dashed line represents the best
fit by Eq.~\eq{eq:fit}.}
\label{fig:result}
\end{figure}

In Fig.~\ref{fig:result} we show the dimensionless coefficient~\eq{eq:coeff} of
the conductivity~\eq{eq:sigma} as a function of temperature~$T$. In agreement
with our previous results~\cite{Braguta:2013loa}, the dissipationless energy
transfer is absent in the confinement phase. The conductivity coefficient
$C_{\mathrm{AME}}(T)$ raises with temperature at phase transition region, and
approaches a constant value at $T \sim 500 \, {\mathrm{MeV}}$ [$T \sim 1.5 \,
T_c$ 
for the SU(2) gauge theory] implying the $T^2$ behavior
of the conductivity $\sigma(T)$ at higher temperatures.

We find the the temperature behavior of the coefficient $C_{\mathrm{AME}}$ can
well be described by the following function,
\beqn
C^{\mathrm{fit}}_{\mathrm{AME}}(T) = C^{\infty}_{\mathrm{AME}} \, \exp \left( -
\frac{h \, T_0}{T - T_0} \right)\,,
\quad T> T_0\,, \quad
\label{eq:fit}
\eeqn
with the best fit parameters
\beqn
C_{\mathrm{AME}}^\infty = 0.0097(2)\,,
\label{eq:C:AME:inf}
\eeqn
$h = 0.055(7)$ and $T_0 = 339(2)\,{\mathrm{MeV}}$.  The best fit value for the
temperature scale is quite close to the pseudocritical temperature of the
deconfinement transition of the lattice $SU(2)$ gauge theory at our lattices. 
The quality of the fit~\eq{eq:fit} is given by
$\chi/{\mathrm{d.o.f.}} = 1.8$. The fit is shown in Fig.~\ref{fig:result} by the
dashed line.

The quantity~\eq{eq:C:AME:inf} corresponds to the AME conductivity $\sigma(T) =
C_{\mathrm{AME}}^\infty \, T^2$ in the high-temperature limit.  For a conformal
theory with two colors of fermions and one single flavor, the theoretical
proportionality coefficient should be an order of magnitude
larger~\eq{eq:sigma:theor}, $C^{\mathrm{th}}_{\mathrm{AME}} = 1/6 \approx
0.166$. The ratio between the observed and predicted coefficients is the same as
in our previous study~\eq{eq:ratio}. Notice that in lattice simulations of free fermions 
the anomalous energy flow agrees very well with the theoretical 
predictions~\cite{Buividovich:2013hza} 
while we observe a large discrepancy between theoretical and numerical results in the 
lattice simulations of the interacting gauge theory.

Figure~\ref{fig:result} also demonstrates the robustness of the result with
respect to variations of the lattice volume. For example, the results at 
$T \approx 400\, {\mathrm{MeV}} \approx 1.3\, T_c$) and 
$T \approx 720\, {\mathrm{MeV}} \approx 2.37\, T_c$ stay unchanged within the
error bars as the
volume changes in the range $V = (8 \dots 15)\,{\mathrm{fm}}^3$ and $V = (1.3
\dots 2.6)\,{\mathrm{fm}}^3$, respectively. This behavior is contrasted with the
simulations of the same effect in a theory with free fermions, where large
finite-volume corrections were observed~\cite{Buividovich:2013hza}. 
The energy flow is almost insensitive to variations of the ultraviolet 
cutoff given by inverse lattice spacing~\cite{Braguta:2013loa}.

Concluding, we have numerically calculated the temperature behavior of the
dissipationless energy flow induced by the background axial magnetic field (the
Axial Magnetic Effect) in the quenched lattice $SU(2)$ gauge theory. We show
that the energy flow is absent in the confinement phase. In the deconfinement
phase the conductivity flow is proportional to the strength of the axial
magnetic field. The AME conductivity raises sharply in the phase transition
region at $T \sim T_c$ and reaches the expected $T^2$ behavior as the
temperature increases over $1.5 \, T_c$. However, the numerically found conductivity 
coefficient is approximately 17 times smaller than the
coefficient predicted by the linear response theory at weak coupling. 

\acknowledgments

We thank P.~V.~Buividovich for useful discussions.
The work of the Moscow group was supported by Grants 
RFBR-13-02-01387, RFBR-14-02-01185, MD-3215.2014.2, Federal Special-Purpose
Program "Cadres" of the
Russian Ministry of Science and Education, and by a grant from the
FAIR-Russia Research Center. Numerical calculations were performed at
the ITEP computer systems ``Graphyn'' and ``Stakan''. The work of M.N.C.
was partially supported by grant No. ANR-10-JCJC-0408 HYPERMAG of
Agence Nationale de la Recherche (France). The work of K.L. was
partially supported by by Plan Nacional de Altas Energ\'\i FPA2012-32828, 
Consolider Ingenio 2010 CPAN (CSD2007-00042); HEP-HACOS S2009/ESP-1473,
SEV-2012-0249. This project is also supported by the Far Eastern Federal University, grant 13-09-617-m\_a.

 \end{document}